\documentstyle[12pt]{article}

\begin{document}
\begin{center}
STATIONARY BLACK HOLES IN A GENERALIZED
THREE-DIMENSIONAL THEORY OF GRAVITY \\
\vskip 5mm
{\bf Paulo M. S\'a} \\
\vskip 0.2cm
{\scriptsize Sector de F\'{\i}sica,
	     Unidade de Ci\^encias Exactas e Humanas,
	     Universidade do Algarve,} \\
{\scriptsize Campus de Gambelas,
	     8000 Faro, Portugal}.
\vskip 0.4cm
{\bf Jos\'e P. S. Lemos} \\
\vskip 0.2cm
{\scriptsize  Departamento de Astrof\'{\i}sica,
	      Observat\' orio Nacional-CNPq,} \\
{\scriptsize  Rua General Jos\'e Cristino 77,
	      20921 Rio de Janeiro, Brasil,} \\
{\scriptsize  \&} \\
{\scriptsize  Departamento de F\'{\i}sica,
	      Instituto Superior T\'ecnico,} \\
{\scriptsize  Av. Rovisco Pais 1, 1096 Lisboa, Portugal.} \\
\end{center} 
 
\medskip

\begin{abstract}
\noindent
We consider a generalized three-dimensional theory of gravity
which is specified by two fields,
the graviton and the dilaton,
and one parameter.
This theory contains, as particular cases,
three-dimensional General Relativity
and three-dimensional String Theory.
Stationary black hole solutions are generated from the
static ones using a simple coordinate transformation.
The stationary black holes solutions thus obtained are locally
equivalent to the corresponding static ones,
but globally distinct.
The mass and angular momentum of the stationary black hole
solutions are computed using an extension of the
Regge and Teitelboim for\-ma\-lism.
The causal structure of the black holes is described.
\end{abstract}

\newpage

\noindent
{\bf 1. Introduction}

\vskip 3mm

A few years ago, Ba\~nados, Teitelboim and Zanelli (BTZ)
showed that three-dimensional General Relativity with a negative
cosmological constant admits a black hole \cite{btz}.
This solution exhibits many of the characteristics of four
dimensional black holes:
it arises from the gravitational collapse of matter
\cite{mann},
it is a thermodynamic object with a well defined temperature
and entropy \cite{brow}
and it can be matched to a three-dimensional perfect fluid star
\cite{cruz}.
Furthermore, the BTZ black hole can be transformed into a
four-dimensional black string in General Relativity \cite{zanc}
(for a review see \cite{carl,mann2}).

A interesting aspect of the BTZ black hole
lies in its relation with string theory.
This black hole was shown to be also a solution to low energy
string theory with a non-vanishing antisymmetric tensor
\cite{horo,kalo}.
Under duality it is equivalent \cite{horo}
to a black string \cite{horn} which,
in the non-rotating case,
is simply the three-dimensional black hole obtained by taking
the product of the two-dimensional string black hole
\cite{mand,witt} with $S^1$.
It was further shown by Kaloper \cite{kalo}
that the BTZ black hole is also a solution of topologically
massive gravity.
Thus, the BTZ black hole is a solution of alternative
theories of gravity.
That fact has attracted considerable interest
to the study of black objects in
three-dimensional theories of gravity
and led to some generalizations of the BTZ black hole.

Of particular interest are the generalizations of the BTZ
black hole which are achieved by the introduction of a dilaton
field \cite{chan,skl}.
One such generalization can be acomplished by considering 
a action of the Brans-Dicke type,
\begin{equation}
S=\frac{1}{2\pi} \int d^3x \sqrt{-g} e^{-2\phi}
  \left[ R - 4 \omega \left( \partial \phi \right)^2 
	   + 4 \lambda^2 
  \right],                         \label{eq:1}
\end{equation}
where
$g$ is the determinant of the three-dimensional metric,
$R$ is the curvature scalar,
$\phi$ is a scalar field,
$\lambda$ is a constant and $\omega$ is the three-dimensional
Brans-Dicke parameter.
Each different $\omega$ yields a different dilaton gravity theory;
$\omega=-1$ corresponds to the low energy string action with
vanishing antisymmetric tensor field,
$\omega=0$ corresponds to a theory related to four-dimensional
General Relativity with one Killing vector and
for $\omega=\pm\infty$ one obtains three-dimensional General Relativity.
Despite its simplicity, the dilaton gravity theory given by the
above action admits a very rich structure of black hole
solutions.

In a previous work \cite{skl}, we have found and analysed,
for the full range of values of the parameter $\omega$,
the static black hole solutions obtained from (\ref{eq:1})
with the following {\it ansatz} for the metric:
\begin{equation}
     ds^2 = - e^{2\nu(r)} dt^2 + e^{-2\nu(r)}dr^2 + r^2 d\varphi^2,
	      \quad\quad 0\leq\varphi\leq 2\pi.
			       \label{eq:4}
\end{equation}
These black hole solutions have the form:
\begin{eqnarray}
 ds^2 &=& -\left[
	    a^2r^2 - \frac{b}{(ar)^{\frac{1}{\omega+1}}}
	    \right] dt^2
	   + \frac{dr^2}{a^2r^2 - \frac{b}{(ar)^{\frac{1}{\omega+1}}}}
	   + r^2 d\varphi^2,                      \nonumber \\
      & &  \hskip 5cm \omega \neq -\frac32,-1,
					     \label{eq:10}  \\
 ds^2 &=&    4\lambda^2 r^2 \ln (br) dt^2
	   - \frac{dr^2}{4\lambda^2 r^2 \ln (br)}
	   + r^2 d\varphi^2,                      \nonumber \\
      & &  \hskip 5cm \omega=-\frac32,
					     \label{eq:12}
\end{eqnarray}
with the dilaton field given by:
\begin{eqnarray}
 e^{-2\phi} &=& (ar)^{\frac{1}{\omega+1}},   
	   \hskip 0.8cm \omega \neq-1.         \label{eq:9a}
\end{eqnarray}
The constant of integration $b$ is positive and related to 
the mass of the black holes,
\begin{eqnarray}
& & M=\frac{\omega+2}{\omega+1}b, \hskip 31mm
	 \omega \neq-\frac32,-1, \\
& & M=-4\lambda^2 \ln (b), \hskip 25mm \omega=-\frac32;
\end{eqnarray}
the constant $a$ is related to $\lambda$ and $\omega$.

For $\omega=-1$,
which corresponds to the low energy three-dimensional string
theory with a vanishing antisymmetric tensor field,
the metric gives simply the three-dimensional Minkowski spacetime
and the dilaton is constant.
Thus, in what follows the case $\omega=-1$ will not be considered.

The causal structure, the geodesic motion of null and timelike
particles in the black hole geometries and the ADM masses of the
black hole solutions (\ref{eq:10})-(\ref{eq:9a}) were analysed
in detail in ref.~\cite{skl}.

Recently, Cadoni \cite{cado} has analysed, from the point of
view of duality symmetries, the theory given by action (\ref{eq:1})
and recover, in particular, the static solution (\ref{eq:10}).

The purpose of this paper is to generalize the static black hole
solutions (\ref{eq:10})-(\ref{eq:9a}) in order to include rotation.

\vskip 1cm

\noindent
{\bf 2. Stationary Black Hole Solutions}

\vskip 3mm

Stationary black hole solutions of action (\ref{eq:1}) 
can be generated from the static black hole solutions 
by a simple coordinate transformation:
\begin{eqnarray}
  t        &\rightarrow& \beta t -\frac{\theta}{a^2} \varphi, \\
  \varphi  &\rightarrow& \beta \varphi - \theta t,
\end{eqnarray}
where $\beta$, $\theta$ and $a$ are constants.
The stationary black hole solutions thus obtained are locally
equivalent to the corresponding static ones,
but globally distinct for each value of the
parameter $\theta$ \cite{stac}.

Applying the above coordinate transformation
to solutions (\ref{eq:10}) and (\ref{eq:12})
one obtains the following stationary black holes
(for $\omega=-\frac32$ we make $a^2=\lambda^2$):
\begin{eqnarray}
 ds^2 &=& -\left[
	    \left( \beta^2-\frac{\theta^2}{a^2} \right)
	    a^2r^2 - \frac{\beta^2 b}{(ar)^{\frac{1}{\omega+1}}}
	    \right] dt^2
	  -\frac{\beta \theta}{a^2}\frac{b}{(ar)^{\frac{1}{\omega+1}}}
	   2d\varphi dt      \nonumber \\
      & & + \frac{dr^2}{a^2r^2 - \frac{b}{(ar)^{\frac{1}{\omega+1}}}}
	  + \left[ \left(\beta^2-\frac{\theta^2}{a^2} \right) r^2 
	      +\frac{\theta^2}{a^4} \frac{b}{(ar)^{\frac{1}{\omega+1}}}
	      \right] d\varphi^2,                      \nonumber \\
      & &  \hskip 5cm \omega \neq -\frac32,-1,
					     \label{eq:102}  \\
 ds^2 &=&    \left[ 4\lambda^2 \beta^2 r^2 \ln (br)
		   + \theta^2 r^2 \right] dt^2
	   - \left[ 4 \beta \theta r^2 \ln (br)
		   + \beta \theta r^2 \right] 2d\varphi dt \nonumber \\
      & &  - \frac{dr^2}{4\lambda^2 r^2 \ln (br)}
	   + \left[ \frac{4\theta^2}{\lambda^2} r^2 \ln (br) 
		   + \beta^2 r^2 \right] d\varphi^2, \nonumber \\
      & &  \hskip 5cm \omega=-\frac32.
					     \label{eq:103} 
\end{eqnarray}
The dilaton field is given by eq.~(\ref{eq:9a}).

For the case $\omega \neq -\frac32,-1$,
in order to have the standard form of the anti-de Sitter spacetime
at spatial infinity we choose $\beta^2=1+\frac{\theta^2}{a^2}$.
For $\omega=-\frac32$ we choose $\beta=1$.

The $\omega=0$ stationary black hole
was discussed in ref.~\cite{lemo}, 
where the three-dimensional gravity theory
was obtained through dimensional reduction from four-dimensional
General Relativity with one Killing vector field.
This solution was further generalized to include
charge \cite{zanc2}.
Some of these black hole solutions were also analysed
in ref.~\cite{chan2}; the action considered there differs from
our action by a conformal transformation 
$g_{ij}\rightarrow e^{-4\phi} g_{ij}$.

To define mass and angular momentum for
the black hole solutions (\ref{eq:102})-(\ref{eq:103}) 
we apply the formalism of Regge and Teitelboim \cite{regg}
(see also \cite{skl,lemo}).

We write the stationary metric in the canonical form,
\begin{equation}
     ds^2 = - (N^0)^2 dt^2
	    + \frac{dr^2}{f^2}
	    + H^2 \left(N^{\varphi}dt +d\varphi \right)^2,
			       \label{eq:104}
\end{equation}
where $N^0(r)$ is the lapse function, 
$N^{\varphi}(r)$ is the shift function,
$f(r)$ and $H(r)$ are some functions of $r$.

Then,
the Hamiltonian form of action (\ref{eq:1})
can be written as:
\begin{eqnarray}
 S &=& -\Delta t \int dr N
	\left[ \frac{2\Pi^2}{H^3} e^{-2\phi}
	      -4f^2 (H \phi_{,r} e^{-2\phi})_{,r} 
	      -2H\phi_{,r} (f^2)_{,r} e^{-2\phi} \right. \nonumber \\
 & &  \ \ \ \ \ \ \ \ \ \ \ 
      + \left. 2f(fH_{,r})_{,r} e^{-2\phi}
      + 4\omega H f^2 (\phi_{,r})^2 e^{-2\phi}
      - 4\lambda^2 H e^{-2\phi} \right] \nonumber \\
 & &  + \Delta t \int dr N^{\varphi} 
	\left[ 2\Pi e^{-2\phi} \right]_{,r}
      + B,
			       \label{eq:105}
\end{eqnarray}
where $\Pi^{r\varphi}= -\frac{H N^{\varphi}_{\ ,r}}{2N}$
($\Pi\equiv\Pi^{r}_{\ \varphi}$) is the momentum conjugate to
$g_{r\varphi}$, $N=\frac{N^0}{f}$
and $B$ is a surface term.

Upon varying the action with respect to
$f(r)$, $H(r)$, $\Pi(r)$ and $\phi(r)$
one picks up aditional surface terms,
which have to be canceled by $\delta B$
(in order that Hamilton's equations are satisfied):
\begin{equation}
 \delta B = - \Delta t N \delta M
	    + \Delta t N^{\varphi} \delta J,
\end{equation}
where $M$ and $J$ are, respectively, the mass and the angular
momentum of the black holes:
\begin{eqnarray}
 \delta M &=& (2H\phi_{,r}-H_{,r}) e^{-2\phi} \delta f^2
      - 2H \left[ (f^2)_{,r} + 4(\omega+1) f^2 \phi_{,r} \right]
		e^{-2\phi} \delta \phi
					   \nonumber \\
   & & +(f^2)_{,r} e^{-2\phi} \delta H
       + 4 f^2 H e^{-2\phi} \delta (\phi_{,r})
       - 2f^2 e^{-2\phi} \delta (H_{,r}), 
					   \label{eq:106} \\
 \delta J &=& 4\Pi e^{-2\phi} \delta \phi -2 e^{-2\phi} \delta \Pi.
					   \label{eq:107}
\end{eqnarray}

We choose the zero point of energy in such a way that 
the mass and angular momentum vanish when the horizon size
goes to zero.
This means that the empty space is given by:
\begin{equation}
 ds^2 = - a^2 r^2 dt^2
	+ \frac{dr^2}{a^2 r^2}
	+ r^2 d\varphi^2,
\end{equation}
which asymptotically coincides with the anti-de Sitter spacetime. 
For $\omega=-3/2$ the above described procedure of choosing the
zero point of energy does not apply,
since for any (positive) value of $b$
one still has a black hole solution.
In other words, the black hole spectrum is unbounded and there is
no ground state.
Thus,
for $\omega=-3/2$ the zero point of energy is choosen
(arbitrarily) to correspond to the black hole solution with $b=1$.

Inserting in eqs.~(\ref{eq:106}) and (\ref{eq:107})
the solutions (\ref{eq:9a}), (\ref{eq:102}) and (\ref{eq:103}) 
one obtains for $\omega\neq -\frac32,-1$:
\begin{eqnarray}
 M &=& \left[ \frac{\omega + 2}{\omega +1}
	     +\frac{2\omega+3}{\omega+1} \frac{\theta^2}{a^2}
       \right] b,
       \label{eq:108} \\
 J &=& \frac{2\omega + 3}{\omega +1} 
       \sqrt{1+\frac{\theta^2}{a^2}}
       \frac{\theta}{a^2} b,
       \label{eq:109}
\end{eqnarray}
and for $\omega = -\frac32$:
\begin{eqnarray}
 M &=& -4\lambda^2 \left[  1 + \ln (b)
	   \left( 1-\frac{2\theta^2}{\lambda^2}\right)
	 -\sqrt{1-\frac{4\theta^2}{\lambda^2} \ln (b)}           
		   \right],
       \label{eq:110} \\
 J &=& 0.
       \label{eq:111}
\end{eqnarray}

Let us now analyse the solutions
obtained for $\omega\neq-\frac32,-1$.
We first consider the solutions with positive mass, $M>0$.
Using eqs.~(\ref{eq:108}) and (\ref{eq:109}), the metric
(\ref{eq:102}) can be put in the form:
\begin{eqnarray}
 ds^2 &=& -\left[
	   a^2r^2 - \frac{\omega+1}{2(\omega+2)}
		\frac{M(1+\Omega)}{(ar)^{\frac{1}{\omega+1}}}
	   \right]
	   dt^2
	  -\frac{\omega+1}{2\omega+3}
	   \frac{J}{(ar)^{\frac{1}{\omega+1}}}
	   2d\varphi dt                                  \nonumber \\
      & & +\frac{dr^2}{a^2r^2 - \frac{1}{2(\omega+2)}
	   \frac{M[(2\omega+3)\Omega-1]}{(ar)^{\frac{1}{\omega+1}}}}
	  +\frac{1}{a^2} \left[
	   a^2r^2 + \frac{M(1-\Omega)}{2(ar)^{\frac{1}{\omega+1}}}
	   \right]
	   d\varphi^2,                                   \nonumber \\
      & &  \hskip 55mm \omega \neq -2, -\frac32, -1, \label{eq:112} \\
 ds^2 &=& -\left( a^2r^2 - \frac{a^2 J^2}{M}ar \right) dt^2
	  -Jar 2d\varphi dt 
	  +\frac{dr^2}{a^2r^2 - M\left( \frac{a^2J^2}{M^2} -1\right) ar}
							 \nonumber \\
      & & +\frac{1}{a^2}
	   \left( a^2r^2 + M ar \right) d\varphi^2,
	   \hskip 20mm \omega = -2.                  \label{eq:112a}
\end{eqnarray}
where we have introduced the notation:
\begin{equation}
  \Omega =\sqrt{1-\frac{4(\omega+1)(\omega+2)}
	     {(2\omega+3)^2}\frac{a^2 J^2}{M^2}}.
	     \label{eq:113}
\end{equation}

The condition that $\Omega$ remains real imposes for
$-\infty<\omega<-2$ and $-1<\omega<+\infty$ a restrition
on the allowed values of the angular momentum:
\begin{equation}
|aJ|\leq \frac{|2\omega+3|M}{2\sqrt{(\omega+1)(\omega+2)}}.
\label{eq:130}
\end{equation}

The metric components become singular at $r=0$.
An inspection of the Ricci and Kretschmann scalars,
\begin{eqnarray}
   & & R =
     - 6a^2
     - \frac{\omega}{(\omega+1)^2}
       \frac{ba^2}{(ar)^{\frac{2\omega+3}{\omega+1}}},
					    \label{115} \\
   & & R_{abcd}R^{abcd} =
       12a^4
     + \frac{4\omega}{(\omega+1)^2}
       \frac{ba^4}{(ar)^{\frac{2\omega+3}{\omega+1}}}
     + \frac{3\omega^2 + 8\omega +6}{(\omega+1)^4}
       \frac{b^2a^4}{(ar)^{\frac{4\omega+6}{\omega+1}}},
					    \label{116}
\end{eqnarray}
where $b$ is given by:
\begin{eqnarray}  
  & & b=M\frac{(2\omega+3)\Omega-1}{2(\omega+2)},
      \hskip 22mm    \omega \neq -2, -\frac32, -1,  \label{eq:114} \\
  & & b=M\left( \frac{a^2J^2}{M^2}-1 \right),
      \hskip 25mm    \omega=-2,                     \label{eq.114a}
\end{eqnarray}
shows that $r=0$ is a curvature singularity 
for $-\frac32>\omega>-1$.
For $-\frac32<\omega<-1$ the curvature singularity
is located at $r=+\infty$ and the
surface $r=0$ corresponds rather to a topological singularity.

Depending on the values of the parameter $\omega$ and the relation 
between the mass and the angular momentum,
the metric components can also become singular 
for another value of the coordinate $r$, namely
$ar=b^{\frac{\omega+1}{2\omega+3}}$,
which corresponds to the horizon.
For $-\infty<\omega<-2$ there is always an horizon
(for the allowed values of $J$ given by eq.~(\ref{eq:130})),
while for $-2<\omega<-\frac32$ such surface never exists,
i.e., there are no black hole solutions with positive mass
for this range of values of $\omega$.
For $-1<\omega<+\infty$ the situation is similar
to the one occurring for the four-dimensional spinning black hole, 
namely, the horizon exists for $|aJ|<M$.
Note, however, that the extreme case $|aJ|=M$ is not
anymore a black hole, but rather a naked singularity.
A rather exotic behaviour occurs for $\omega=-2$ 
and $-\frac32<\omega<-1$;
for these values of $\omega$
the black hole has to spin very fast, $|aJ|>M$;
as soon as the angular momentum decreases, $|aJ|\leq M$,
the solution changes character and becomes a naked singularity.
For $-\frac32<\omega<-1$ there is a topological singularity
at $r=0$.
For this reason the spacetime manifold cannot be extended for
negative values of $r$, contrary to the two-dimensional case
\cite{lesa}.

Table~1 gives a summary of the conditions imposed on the
angular momentum of the positive mass black holes.

For $\omega=\infty$  and $|aJ|<M$,
contrarily to cases mentioned above,
the solution admits two horizons, given by
$ar_{\pm}=\pm\sqrt{M\Omega}$.
Indeed, for $r=0$ there are no curvature or topological
singularities and thus spacetime can be extended for
negative values of $r$.
By performing the coordinate transformation
$r\rightarrow \bar{r}$, where $\bar{r}$ is given by
$\bar{r}^2=r^2 + \frac{M(1-\Omega)}{2a^2}$,
one obtains the usual form of the BTZ black hole metric
with outer and inner horizons.
This solution was extensively analysed in refs.~\cite{bhtz,carl}.
The Penrose diagram can be infinitely
extended, similary to the Penrose diagram of the four-dimensional
Kerr black hole.
The surface $\bar{r}=0$ is timelike and corresponds
to a singularity in the causal structure
and thus the solution cannot be extended for negative values
of $\bar{r}$.
Here, contrary to the case $\omega\neq\infty$,
there is an extreme black hole,
obtained for $|aJ|=M$. 
For $J=0$, the solution admits only one horizon and 
the causal structure is similar to the one exhibited 
by the black hole solutions corresponding
to $-\infty<\omega<-2$ and $-1<\omega<+\infty$,
with the exception that the spacetime surface $\bar{r}=0$
is a Taub-NUT type singularity in the manifold,
rather than a curvature singularity.

For $M>0$ the $g_{\varphi \varphi}$ component of the metrics
(\ref{eq:112}) and (\ref{eq:112a})
is always positive, which implies that there
are no closed timelike curves.

We have been considering only solutions with
positive mass, but in fact there is no reason to exclude the
solutions with negative mass from the physical spectrum,
since they do not correspond to naked singularities,
but rather to black holes.
An analysis identical to the one carried above for the positive
mass black holes can also be performed. In table~1 we summarize
the conditions that angular momentum must obey in order that the
negative mass solutions correspond to black holes.
We also remark that these solutions admit closed timelike curves.
Indeed, the coordinate $\varphi$ becomes timelike in a certain
region of spacetime. Since this coordinate is periodic one concludes
that there are closed timelike curves in this region of spacetime.

Let us now turn to the solution
obtained for $\omega=-\frac32$ (see eq.~(\ref{eq:103})).

The stricking feature of this solution is that both
surfaces $r=0$ and $r=\infty$ correspond to a curvature
singularity. Indeed, the Ricci and Kretschmann scalars,
\begin{eqnarray}
& & R = 4\lambda^2 \left[ 5+6\ln (br) \right],
\\
& & R_{abcd}R^{abcd} =
16\lambda^2 \left[ 11+20\ln (br) +12 \ln^2 (br) \right],
\end{eqnarray}
diverge for $r=0$ and $r=\infty$.

The condition that $M$ remains real imposes that the constant
$b$ has to be less or equal to $e^{\frac{\lambda^2}{4\theta^2}}$
(see eq.~(\ref{eq:110})).
For any value of the constant $b$ in the interval
$0<b\leq e^{\frac{\lambda^2}{4\theta^2}}$,
the solution has an horizon, given by $br=1$.
The zero point of energy was chosen (see above),
so that the black hole has zero mass for $b=1$.
This choice of the ground state divides arbitrarily 
the mass spectrum in states with positive and negative
masses.       
Note that although there is dragging of inertial frames,
the angular momentum of the solution is always zero
(see eq.~(\ref{eq:111})).
This kind of behaviour is unexpected
for a stationary solution.
The $g_{\varphi\varphi}$ component of the metrics
(\ref{eq:103}) becomes negative
for $r<r_{CTC}=\frac{1}{b} \exp \{ -\frac{\lambda^2}{4\theta^2} \}$.
Thus,
there are closed timelike curves in the region $0<r<r_{CTC}$.

The solution for $\omega=-\frac32$ is thus a
non-spinning black hole with closed timelike curves.

The Penrose diagrams of the stationary black hole solutions described
above are identical to the Penrose diagrams of the static black holes
(see ref.~\cite{skl}), so they will not be reproduced here.

\vskip 5mm

\noindent {\bf Acknowledgments:}

\noindent
This work was supported in part by {\it Junta Nacional de
Investiga\c c\~ao Cient\' \i fica e Tecnol\'ogica}, Portugal.

\vfill
\newpage
\vskip 5mm

\noindent
\begin{tabular}{|l|l|l|}
\hline
Range of $\omega$           &
Black holes with $M>0$      &
Black holes with $M<0$      \\ \hline\hline
$-\infty<\omega<-2$         & 
$|aJ|\leq\frac{|2\omega+3|M}{2\sqrt{(\omega+1)(\omega+2)}}$  &
never                       \\ \hline
$\omega=-2$                 &
$|aJ|>M$                    &
$|aJ|<|M|$                  \\ \hline
$-2<\omega<-\frac32$        &
never                       &
always                      \\ \hline
$\omega=-\frac32$           &
$J=0$                       &
$J=0$                       \\ \hline
$-\frac32<\omega<-1$        &
$|aJ|>M$                    &
$|aJ|<|M|$                  \\ \hline
$-1<\omega<+\infty$         &
$|aJ|<M$                    &
$|M|<|aJ|\leq \frac{(2\omega+3)|M|}{2\sqrt{(\omega+1)(\omega+2)}}$ 
			    \\ \hline
\end{tabular}

\vskip 10mm

\noindent
{\bf Table Caption}
\vskip 5mm
\noindent
{\bf Table 1.}

\noindent
Values of the angular momentum for which the black holes
have positive and negative masses.

\end{document}